\documentclass[11pt,a4paper]{article}
\usepackage{jheppub}
\usepackage{graphicx}

\def\hbar{\hspace{0pt}\raisebox{1pt}{$-$} \hspace{-7pt} h}

\def\5{\overline 5}

\newcommand{\ba}{\begin{eqnarray}}
\newcommand{\ea}{\end{eqnarray}}
\newcommand{\no}{\nonumber}
\newcommand{\be}{\begin{equation}}
\newcommand{\ee}{\end{equation}}
\newcommand{\bea}{\begin{eqnarray}}
\newcommand{\eea}{\end{eqnarray}}

\def\NG{Nambu-Goldstone }

\def\SM{standard model~}

\def\SDE{Schwinger-Dyson equation }
\def\SD{Schwinger-Dyson }
\title{Technicolor at Criticality}
\date{\today
}
\author{
Luca Vecchi}
\affiliation{Theoretical Division T-2, Los Alamos National Laboratory \\ Los Alamos, NM 87545, USA}
\emailAdd{vecchi@lanl.gov}
\abstract{

We discuss an asymptotically \textit{non-free}, natural model for dynamical electro-weak symmetry breaking characterized by the emergence of a weakly coupled Higgs in the IR regime. Due to the large anomalous dimension of the Higgs operator, the model is capable of solving the hierarchy problem without losing the phenomenologically appealing features typical of weakly coupled Higgs sectors.

We speculate on the possibility that such a scenario be realized as a strongly coupled phase of non-supersymmetric non-abelian gauge theories.

}

\begin{document}
\maketitle

\section{Introduction}

Asymptotically free non-abelian gauge theories provide the most elegant solution to the hierarchy problem. Technicolor (TC) is the simplest realization of this idea~\cite{TC}.


Non-abelian gauge theories solve the hierarchy problem because they have no strictly relevant (as opposed to marginally relevant) deformation in their UV formulation~\footnote{This last property is at the root of naturalness. Asymptotic freedom has nothing to do with the solution of the hierarchy problem.}: the IR scale $\Lambda_\chi$ generated by dimensional transmutation in asymptotically free TC depends on the short distance physics only very mildly, and the theory naturally accounts for a separation, \textit{and} stabilization between the electro-weak scale $v=250$ GeV and the UV cutoff. 

In order for a Higgs sector to be natural, it is crucial that its UV dynamics has no weakly coupled scalar particles -- in particular that there is no strictly relevant UV mass term for the Higgs. Yet, the absence of a weakly coupled Higgs field in the IR regime appears to be the ultimate source of most of the phenomenological issues plaguing models for dynamical electro-weak (EW) symmetry breaking, namely fermion mass generation and precision measurements.

In models with dynamical EW symmetry breaking, the generation of fermion masses requires the introduction of mediators between the \SM fermions $\psi$ and the techni-quark bilinear $\bar QQ$, which plays the role of the EW order parameter in these models. This mediation is usually achieved with an extended technicolor sector~\cite{ETC}, an effective description of which may be formally given in terms of a set of 4-fermion contact terms with structures $(\bar QQ)^2$, $\bar QQ\psi\psi$, and $(\psi\psi)^2$.
The scale $\Lambda_{ETC}$ suppressing these operators should be sufficiently large in order to avoid large FCNC effects induced by the 4-$\psi$'s operators, but it cannot be too large if we are willing to generate realistic masses for the standard model (SM) fermions.

Following~\cite{CTC}, one can find an upper bound for the flavor violation scale $\Lambda_{ETC}$ by estimating the energy at which the SM fermions, in particular the top quark, become strongly coupled to the Higgs sector. This scale is a strong function of the scaling dimension $\Delta_{\bar QQ}$ of the Higgs operator such that $\Lambda_{ETC}\rightarrow\infty$ as the scaling dimension approaches the value $\Delta_{\bar QQ}=1$ typical of a weakly coupled scalar.

In an attempt to alleviate the flavor problem in models for dynamical EW symmetry breaking, Holdom suggested to consider asymptotically \textit{non-free} scenarios~\cite{Holdom}. Asymptotically non-free theories remain strongly coupled in the UV ($\mu>\Lambda_\chi$), and may induce large anomalous dimensions for the techni-quark bilinear. The tension emerging from the requirement that a high flavor scale $\Lambda_{ETC}$ be compatible with the large top mass could be considerably relaxed in a strong dynamics in which the scaling dimension of the techni-quark bilinear can be pushed close to the value $\Delta_{\bar QQ}=1$. In walking technicolor (WTC)~\cite{Holdom}\cite{WTC}, for instance, one roughly expects $\Delta_{\bar QQ}\gtrsim2$ in the whole range $\Lambda_\chi\lesssim\mu\lesssim\Lambda_{ETC}$. This translates into a strong coupling scale for the top Yukawa of order $\sim4\pi\Lambda_\chi$~\cite{CTC}, a somewhat larger energy than expected in a QCD-like TC.

In addition to alleviating the flavor problem, asymptotically non-free models may reduce the tension with the EW precision measurements that models of dynamical breaking have to face. In asymptotically non-free models the perturbative estimates, or even the QCD-rescaled estimates, of the Peskin-Takeuchi S parameter~\cite{S} are certainly inadequate, and one cannot use those arguments to rule out such scenarios. Moreover, in an asymptotically non-free dynamics it is in principle possible to arrange the condition $\Delta_{\bar QQ}<3$; for such scaling dimensions the second Weinberg sum rule is not satisfied, and this is expected to cause a reduction of the S parameter as opposed to asymptotically free TC models in which $\Delta_{\bar QQ}=3$ in the far UV~\cite{Hsu}\cite{SanninoApp}. 

The bottom line of the above discussion is that a model for EW symmetry breaking addressing the hierarchy problem, and capable of pushing the flavor physics far above the weak scale, should posses a somewhat weakly coupled Higgs sector in the IR, say at scales $\mu\sim\Lambda_\chi$, but no weakly coupled scalars in the UV, say for scales $\mu\gg\Lambda_\chi$. Such a scenario requires large anomalous dimensions for the Higgs sector, and ultimately a strong dynamics. 

The conformal TC paradigm proposed in~\cite{CTC} aims precisely to the realization of such a framework. In that model the Higgs dynamics is nearly conformal up to scales of order $\Lambda_{ETC}\gg\Lambda_\chi$ and it is required to satisfy both $\Delta_{\bar QQ}\sim1$ \emph{and} $\Delta_{(\bar QQ)^2}\gtrsim4$ -- where $\Delta_{(\bar QQ)^2}$ denotes the scaling dimension of the Higgs mass operator. The latter condition ensures that the Higgs mass term is not a relevant operator, and hence that the model does not suffer from any naturalness problem. The former condition ensures that the flavor problem can be decoupled to scales compatible with FCNC effects. While these conditions are in principle realizable, calculability still represents the main obstacle towards an explicit realization of this program.

Here we discuss a natural model for EW symmetry breaking in which conformality is badly broken above the weak scale. The model is \emph{large $N$ calculable} and asymptotically \textit{non-free}, and it has an EW order parameter $H=\bar QQ$ with scaling dimension $\Delta_{\bar QQ}$ in the range
\ba\label{gamma}
1\lesssim\Delta_{\bar QQ}\leq2.
\ea
We decided to call this theory \emph{technicolor at criticality} for reasons that will be clarified later. The condition $\Delta_{\bar QQ}=2$ is satisfied in the UV, and implies that the Higgs mass term is a marginally relevant operator at the UV, non-trivial fixed point. This property ensures that the IR physics is only logarithmically sensitive to the UV cutoff, as we will see, and hence that the theory is technically natural. The condition $\Delta_{\bar QQ}\sim1$ will be found to hold in the IR, and it is equivalent to the statement that the Higgs operator in TC at criticality is \emph{weakly coupled}.

Our solution of the naturalness problem differs from the one proposed in~\cite{CTC}. TC at criticality will be shown to be a natural theory for dynamical EW symmetry breaking despite the fact that the large $N$ expansion adopted here forces the relation $\Delta_{(\bar QQ)^2}=2\Delta_{\bar QQ}\leq4$, which would naively indicate a power-law sensitivity of the IR physics on the UV cutoff. We will see in Section~\ref{nat} that large $N$ field theories satisfying~(\ref{gamma}) can be natural provided their dynamics departs rather quickly from IR conformality. This in turn implies that in these models the flavor issue must be addressed at a somewhat lower scale compared to~\cite{CTC}. For the specific case of TC at criticality, we will show that the top physics remains perturbative up to $\sim(150\div200)\times\Lambda_\chi$, which is certainly a significant improvement compared to a WTC scenario. 

The main advantage of our model over that of~\cite{CTC} is that our framework is tractable, at least in principle, within the planar limit. The aim of the present paper is precisely to show that a class of tractable, natural models satisfying~(\ref{gamma}) exists, and accordingly that the very existence of a weakly coupled Higgs boson in the IR \emph{is not} necessarily in conflict with naturalness.

The paper is organized as follows. In Section~\ref{nat} we will discuss the naturalness problem in non-CFT models with IR weakly coupled scalars. We will argue that an IR weakly coupled Higgs sector can evade power-law sensitivity on the UV cutoff if the departure from the condition $\Delta_{\bar QQ}\sim1$ is sufficiently fast as the RG scale increases. A logarithmic running for the Higgs scaling dimension $\Delta_{\bar QQ}$ will be shown to suffice. 

In Section~\ref{TCC} we will present a path integral formulation for technicolor at criticality (TCC), and prove that the model features the property~(\ref{gamma}). We will see that in TCC the departure from the IR condition $\Delta_{\bar QQ}\sim1$ is indeed logarithmic, and that the sensitivity of the IR physics on the UV cutoff is at most logarithmic. 
Here we will also discuss the effect of the large anomalous dimension of the EW order parameter on flavor physics, and interpret the IR condition $\Delta_{\bar QQ}\sim1$ as an indication that the physical Higgs boson in TCC is a pseudo-dilaton of an approximate conformal symmetry. 

In Section~\ref{conj} we will conjecture TCC to be the large $N$ dual, effective description of a strong, asymptotically non-free phase of non-supersymmetric non-abelian gauge theories.

In Section~\ref{Con} we will present our conclusions.

\section{Naturalness and weakly coupled scalars\label{nat}}

In this section we would like to address the following question: How can the condition~(\ref{gamma}) be compatible with naturalness in a large $N$ field theory? Even though our strong Higgs sector is natural when considered in isolation
, it must eventually couple to the SM in order to be a realistic theory for dynamical EW symmetry breaking. Now, since the IR condition $\Delta_{\bar QQ}\sim1$ is expected to receive negligible corrections from the SM interactions: Why does the Higgs mass operator -- which in a large $N$ dynamics as the one considered here would have an IR dimension $\Delta_{(\bar QQ)^2}\sim2$ -- not receive too-large quantum corrections from the SM physics?


The point is that the notion of relevance/irrelevance of an operator is rigorous only in the vicinity of a CFT. In truly weakly coupled Higgs sectors, where conformality is broken by perturbative physics, the scaling dimension $\Delta_{(\bar QQ)^2}\simeq2\Delta_{\bar QQ}$ stays close to 2 for a large energy range. This means that truly weakly coupled Higgs sectors are described by approximate conformal field theories with strongly relevant operators, and are therefore unnatural. In a strong dynamics, on the other hand, the departure from conformality as the RG scale evolves can be rather quick, and one should be a bit more careful. 

In complete generality, assume that the Higgs operator in our large $N$ field theory has an engineering, classical dimension $\Delta_H^{cl}$ and running scaling dimension $\Delta_H$. Focussing for simplicity on the leading order in the planar expansion we take the scaling dimension of the Higgs mass operator to be $2\Delta_H$. The 1-loop radiative corrections to the coupling of the Higgs mass operator induced by the top Yukawa coupling $\bar y_t$ appear at distances $O(1/\Lambda_\chi)$ in the form (see for example~\cite{Ratt}) 
\ba\label{pl}
\sim&&\frac{N_c}{16\pi^2}\,\bar y_t^2(\Lambda)\,\Lambda^{4-2\Delta_H^{cl}}
 \exp\left[-2\int^\Lambda_{\Lambda_\chi} \frac{d\mu}{\mu}\,(\Delta_H-\Delta_H^{cl})\right]\\\no
 =&&\frac{N_c}{16\pi^2}\,\bar y_t^2(\Lambda)\,\Lambda_\chi^{4-2\Delta_H^{cl}}
 \exp\left[\int^\Lambda_{\Lambda_\chi} \frac{d\mu}{\mu}\,(4-2\Delta_H)\right],
\ea
with $\bar y_t(\Lambda)$ the dimensionless running Yukawa coupling evaluated at the UV cutoff $\Lambda$. The exponential term in the first line of~(\ref{pl}) accounts for the RG evolution of the Higgs mass operator from the scale $\Lambda$ down to $\Lambda_\chi$, and $\Delta_H-\Delta_H^{cl}$ is the anomalous dimension of the Higgs field. The UV cutoff dependence given in eq.~(\ref{pl}) reduces to the one found in~\cite{Ratt} in the limit in which $\Delta_H$ is constant, but it also applies to theories in which the RG flow of $\Delta_H$ is not negligible. This latter case will be the focus of our discussion.

Notice that for $\Delta_H\neq1$ the Yukawa couplings $\bar y$ have a nontrivial running already at leading order in the SM couplings. To see this, observe that at leading order in the Yukawa and gauge couplings the Yukawa vertex is not renormalized, and hence the only corrections to $\bar y$ arise from the wave-function renormalization of the Higgs operator. This implies that 
\ba\label{Ybeta}
\mu\frac{d\bar y}{d\mu}=\left[\Delta_H(\mu)-1\right]\bar y+\dots,
\ea
where the dots refer to higher order terms in the couplings between the SM and the Higgs sector. This result says that for $\Delta_H>1$ the Yukawa couplings are irrelevant, and therefore they grow in the UV. This fact will have important implications in what follows.


If the Higgs sector is truly weakly coupled one has $\Delta_H(\mu\lesssim\Lambda)\simeq1$, and from~(\ref{pl}) one recovers the well known fact that the Higgs mass term in a weakly coupled theory is quadratically sensitive to the cutoff. From~(\ref{Ybeta}) it follows that the $\bar y$'s run logarithmically, and the flavor physics stays perturbative up to a very high scale. In the fundamental Higgs model new physics is therefore required to address the hierarchy problem, but there is no hint of an underlying scale of flavor. 

If $\Delta_H(\mu\lesssim\Lambda)=2$ one finds at most a logarithmic dependence on the cutoff from~(\ref{pl}) -- not included in the above formula for brevity -- and concludes that the Higgs sector is natural. This is expected to happen in a WTC framework. In this latter model the scaling dimension of the order parameter $H=\bar QQ$ would be nearly constant, say $\Delta_H\gtrsim2$, in the range $\Lambda_\chi<\mu<\Lambda_{ETC}$, and the Yukawa couplings would scale approximately as
\ba\label{Y}
\bar y(\mu)\sim\bar y(\Lambda_\chi)\left(\frac{\mu}{\Lambda_\chi}\right)^{\Delta_H-1}.
\ea
For the top quark $\bar y_t(\Lambda_\chi)\sim1$, and the top physics becomes strong at a scale $\lesssim4\pi\Lambda_\chi$. Our description breaks down there, and new interactions involving the top quark must become relevant -- the dots in~(\ref{Ybeta}) can no more be neglected. In WTC new physics is therefore required at energies below $\sim4\pi\Lambda_\chi$ to address the flavor problem.

More generally, we would like to see now under which conditions the radiative correction~(\ref{pl}) is compatible with naturalness. An inspection of~(\ref{pl}) reveals that a sufficient condition to have \emph{at most} a logarithmic dependence $\propto(\log\Lambda)^{2\kappa}$ (with $\kappa>0$) on the UV cutoff is 
\ba\label{c}
\Delta_H(\mu)\geq 2 - \frac{\kappa}{\log\mu/\Lambda_\chi}.
\ea
If this condition is satisfied the Higgs mass operator does not receive power-law corrections from scales in the momentum shell $\Lambda_\chi\lesssim\mu\leq\Lambda$. It is hence evident from~(\ref{c}) that the IR relation $\Delta_H\leq2$ in a large $N$ dynamics does not necessarily imply a strong sensitivity on the UV cutoff: to avoid a naturalness problem in large $N$ models 
satisfying $\Delta_H\sim1$ in the deep IR it is sufficient that the IR relation $\Delta_H\sim1$ be violated \emph{sufficiently fast} as the RG scale increases.


The price to pay for these natural, large $N$ models is that the top Yukawa typically becomes non-perturbative at energies closer to the TeV scale than in the conformal TC scenario, where $\Delta_H\sim1$ is assumed to be preserved up to some very high scale~\cite{CTC}. But this is \emph{not} a naturalness problem, and does not represent a drawback. In fact, a relatively low new physics scale might be an indication that the flavor physics in these models could be within the reach of future collider experiments.

We will see in Section~\ref{TCC} that in TCC the running scaling dimension of the Higgs field $H=\bar QQ$ down to scales $O(\Lambda_\chi)$ is given by $\Delta_{\bar QQ}=2-1/\log\mu/\Lambda_\chi$. 
In this case the strongly coupled Higgs sector has at most a logarithmic sensitivity to the cutoff scale: TCC should be considered a natural model despite the fact that in a leading $1/N$ analysis the order parameter has a scaling dimension within the range~(\ref{gamma}).




\section{\label{TCC}Technicolor at Criticality}

We will now present an explicit, and tractable model for dynamical EW symmetry breaking that manifests the appealing feature discussed in the introduction, and in particular~(\ref{gamma}). The theory, called technicolor at criticality for reasons that will be explained in Section~\ref{conj}, can be seen as the non-abelian version of the quenched QED model of Bardeen et al.~\cite{Bardeen}, and it formally arises as a deformation of the CFT defined at the IR fixed point of an asymptotically free, non-supersymmetric technicolor theory. 

Let us define the TC gauge group to be $SU(N)$ and assign the representation $R$ to $N_f$ massless techni-quarks $Q$. These theories are known to possess a \emph{conformal window}, that is a range in flavor space $N_f^c<N_f<N_f^{af}$ in which the long distance physics is described by a CFT. We assume that $N_f$ is chosen within this window. Because we will work in the 't Hooft limit, where $N\rightarrow\infty$ and $N_f$ is kept fixed, we also choose a representation $R$ such that $N_f^c$ stays finite as $N$ is sent to infinity. In this latter case $N_f$ can be fixed while still preserving IR conformality as $N\rightarrow\infty$. For this reason we invoke higher dimensional representations $R$ for the fermionic degrees of freedom. To be definite, we define $R$ to be a two-index representation
, although the following results will have a more general validity~\footnote{The planar limit with $R$ being the fundamental representation is somewhat problematic as it would require a study in the Veneziano limit ($N_f^c\propto N$), for which the factorization of correlator functions of the flavorful fermion bilinear $\bar QQ$ -- a property that will be essential in what follows --, does not apply.}.

Let us now consider the following path integral:
\ba\label{pert}
\langle e^{i\int f(\bar QQ)^2}\rangle_{CFT},
\ea
where $\langle\dots\rangle_{CFT}$ is defined by the CFT correlators of the technicolor theory at the IR fixed point, $\bar QQ$ is the techni-quark bilinear transforming as a bi-fundamental of the flavor group $SU(N_f)\times SU(N_f)$, and $f$ a coupling.

Remarkably, one can extract a number of rigorous predictions from the theory~(\ref{pert}), see~\cite{Witten}\cite{Rastelli}\cite{Vecchi} and~\cite{Vecchi'}. It turns out that the RG evolution of a theory of the form~(\ref{pert}) is significantly simplified if the CFT is a single-trace theory admitting a planar expansion (this is certainly the case if the CFT in~(\ref{pert}) is associated to the IR limit of an asymptotically-free non-abelian theory), and $\bar QQ$ a single-trace scalar (this is also the case for the present theory)~\cite{Witten}~\footnote{The scaling dimension of the single-trace operator must be in the range $2\leq\Delta<3$ for these preditions to apply~\cite{Vecchi}. This is believed to be the case in the conformal window of nonabelian gauge theories~\cite{CG}. Note that this fact provides further support in favor of the conjecture of Section~\ref{conj}.}. If these conditions are satisfied, indeed, the CFT double-trace deformation $(\bar QQ)^2$ does not renormalize the coupling $\lambda$ of the gauge theory at leading order in the planar limit, and the RG flow of~(\ref{pert}) can be entirely encoded in the beta function of the coupling $f$~\cite{Rastelli}\cite{Vecchi}. By working at leading order in the planar expansion, and generalizing the results of~\cite{Witten}, the authors of~\cite{Rastelli}\cite{Vecchi}\cite{Vecchi'} found that the beta function of $f$ and the scaling dimension of the single-trace operator $\bar QQ$ are given by:
\ba\label{beta}
\beta_{\bar f}&=&-\bar f^2+(2\Delta-4)\bar f\\\no
\Delta_{\bar QQ}&=&\Delta-\bar f,
\ea
where $\Delta$ denotes the scaling dimension of the techni-quark bilinear at the IR fixed point of the non-abelian theory, i.e. in the undeformed ($f=0$) CFT of~(\ref{pert}), whereas $\bar f(\mu)$ is the renormalized, dimensionless coupling associated to $f$~\footnote{Consistency with the large $N$ counting requires $f=O\left(1/N^2\right)$. The renormalized coupling has been rescaled so that $\bar f=O(1)$ for convenience.}. Consistently with the above claims, eqs.~(\ref{beta}) describe the RG flow of both quenched QED and the NJL model (the latter model being treatable in space-time dimensions less than 4)~\cite{Vecchi}.

We emphasize that eqs.~(\ref{beta}) are \emph{exact} implications of the theory~(\ref{pert}) at leading order in the planar limit. These represent the starting point of the following discussion.

\subsection{A natural theory for EW symmetry breaking\label{N}}

Taking advantage of~(\ref{beta}) and the analogy with the NJL model and quenched QED, one expects the theory~(\ref{pert}) to manifest chiral symmetry breaking, namely to develop a fermion condensate $\langle\bar QQ\rangle\neq0$, in the phase $\bar f>0$~\cite{Vecchi}. In this section we propose to consider~(\ref{pert}) as a model for dynamical EW symmetry breaking and analyze the implications of the emerging dynamics. A physical interpretation of the construction~(\ref{pert}) will be presented in Section~\ref{conj}.

The model~(\ref{pert}) becomes a natural theory for EW symmetry breaking when the CFT is such that $\Delta=2$. In this case the double-trace deformation is marginally relevant, and the coupling $\bar f$ runs logarithmically. The analogy between the resulting theory and the Gross-Neveu model (the natural version of the NJL theory for chiral symmetry breaking) or the critical version of quenched QED~\cite{Bardeen} is evident. We thus focus on the interesting limit in which the number of massless flavors $N_f$ in the TC gauge theory is chosen to be close to a \emph{critical} value at which the scaling dimension of the techni-quark bilinear at the IR fixed point is $\Delta=2$. We call the theory~(\ref{pert}) with $\Delta=2$ technicolor at criticality (TCC) for obvious reasons.

Let us first discuss whether $\Delta=2$ is a realistic assumption or not. Shortly after we will analyze the implications of the construction~(\ref{pert}). 

For fermions in the symmetric and antisymmetric two-index representation, and using the rainbow approximation to the \SD equation (corrections beyond the ladder approximation are expected to be small in this case~\cite{ALM}), one infers that the condition $\Delta=2$ is realized when the number of fermions is critical, i.e. $N_f=N_f^c$, where~\cite{Sannino}
\ba
N_f^c=\frac{N}{N\pm2}\frac{83 N^2\pm66N-132}{20N^2\pm15N-30},
\ea
with the upper (lower) sign referring to the symmetric (antisymmetric) representation. The critical value $N_f^c$ is not generally an integer, but approaches physical values for moderately large $N$. This indicates that for any integer $N$ the theory has an amount of fine-tuning measured by how much $N_f$ departs from $N_f^c$, i.e. how much $\Delta$ departs from $\Delta=2$. Yet, such a fine-tuning can be made negligible. For example, in the symmetric representation one sees that for $N=50$ the critical number of flavors is $N_f^c=4-\delta\lesssim4$, with $\delta=O(10^{-3})$. We can then take $N_f=4$, so that $N_f\gtrsim N_f^c$. Using again the \SD equation approach one argues that for physical values of $N_f\gtrsim N_f^c$ the IR dimension $\Delta$ of $\bar QQ$ in the asymptotically free phase can be written as $\Delta=2+O(\sqrt{\delta})\gtrsim2$ (see Section~\ref{conj} for more details). As a consequence, the dynamical mass -- estimated conventionally as the scale at which the coupling $\bar f$ blows up -- in a realistic TCC model reads
\ba\label{LL}
\Lambda_\chi=\Lambda e^{-\frac{1}{\bar f(\Lambda)}\left(1+O\left(\frac{\sqrt{\delta}}{\bar f(\Lambda)}\right)\right)}
\ea
In order for our theory~(\ref{pert}) to be natural, i.e. (nearly) \textit{critical}, it will therefore suffice to have $\sqrt{\delta}<\bar f(\Lambda)$ and a moderately large $N$. The latter requirement in particular justifies our planar expansion. 

Having established that the condition $\Delta=2$ (or nearly so) is realizable in a realistic model, in the following we will assume that the field content of the theory~(\ref{pert}) has been chosen so that $\delta$ is negligible and that the UV boundary conditions imply $\Lambda_\chi=O(1)$ TeV. Notice that as opposed to WTC we are requiring $N_f\geq N_c$, i.e. the asymptotically free TC dynamics must posses an IR fixed point in order for TCC to exist (see Section~\ref{conj} for an interpretation of this requirement).

Next we turn to the implications of our construction. A remarkable consequence of~(\ref{pert}) is that the quantum dimension of the field $\bar QQ$ in the leading $1/N$ analysis is~(\ref{beta})
\ba\label{dim}
\Delta_{\bar QQ}(\mu\gtrsim\Lambda_\chi)=2-\frac{1}{\log\left(\mu/\Lambda_\chi\right)},
\ea
and hence it satisfies $\Delta_{\bar QQ}\leq2$ in the perturbative regime. As discussed in Section~\ref{nat}, the running scaling dimension~(\ref{dim}) leads at most to a logarithmic sensitivity to the UV cutoff. The relation $\Lambda_\chi\ll\Lambda$ following from~(\ref{LL}) is therefore radiatively stable in TCC.

Below the scale $\mu\sim\Lambda_\chi$ at which $\Delta_{\bar QQ}\sim1$ perturbation theory cannot be trusted due to the presence of the IR Landau pole at $\mu=\Lambda_\chi$. In physical terms this expresses the fact that below the scale set by the dynamical mass $\Lambda_\chi$ the scaling dimension $\Delta_{\bar QQ}$, and the coupling $\bar f$ itself, become somewhat ambiguous entities due to decoupling of both flavor and gluon degrees of freedom. Fortunately, we will be mostly interested in the physics at scales $\mu>\Lambda_\chi$. Now, because we already have $\Delta_{\bar QQ}\sim1$ for $\mu$ slightly above $\Lambda_\chi$ -- where our perturbative analysis is believed to be reliable -- it is sensible to expect that 
\ba\label{eps}
\Delta_{\bar QQ}(\mu\sim\Lambda_\chi)=1+\epsilon
\ea
for some $\epsilon<1$. Again, the analogy with the Gross-Neveu model or quenched QED confirms this conclusion. 

We thus claim that the scaling dimension of the composite Higgs $\bar QQ$ in TCC is confined in the range~(\ref{gamma}). 
This implies that the $Q$'s in TCC become more and more strongly coupled at larger distances -- where the chiral symmetry is expected to break down -- : the fermions in TCC are \textit{never} weakly coupled, and the theory is said to be asymptotically non-free~\footnote{The UV fixed point $\bar f=0$ of~(\ref{pert}) is clearly not free.}.

\subsection{Flavor physics}
\label{fla}

As reviewed in the introduction, the strong dynamics responsible for EW symmetry breaking should be ultimately coupled to the SM fermions $\psi$. We effectively describe these extended TC interactions as contact terms of the form (we follow the notation of~\cite{Hill})
\ba\label{ETC}
\alpha_{ab}\,\bar Q T^aQ\bar Q T^bQ\quad\quad\beta_{ab}\,\bar\psi T^a\psi\bar Q T^bQ\quad\quad\gamma_{ab}\,\bar\psi T^a\psi\bar\psi T^b\psi
\ea 
on the top of our theory~(\ref{pert}). In the above expression the matrices $T^a$ stand for the generators of the extended technicolor group (in particular the generators of the strong $SU(N)$ and SM symmetries) and chirality operators, while $\alpha_{ab},\beta_{ab},\gamma_{ab}$ are dimensionful coefficients defined in terms of the flavor scales $\Lambda_{ETC}^i$ -- which in general will depend on a family index $i=1,2,3$. Specifically, the operators associated to the $\beta_{ab}$'s ($=O(\bar y/\Lambda_{ETC}^2$)) will induce the SM fermion masses whereas those associated to the $\gamma_{ab}$'s will induce FCNC effects that we would like to suppress.

The addition of a flavor sector, and in particular of the 4-fermion operators associated to the $\alpha_{ab}$'s, would typically deform the short distance physics of TCC, leading perhaps to a more conventional WTC framework. In order for the results of Section~\ref{N} to apply, we should be able to prove that the 4-fermion operators in~(\ref{ETC}) are not relevant at the weak scale, namely that $(\Lambda_\chi/\Lambda_{ETC}^i)^2\ll1$~\footnote{Note that this latter condition also suffices to claim that the operators associated to the $\alpha_{ab}$'s do not spoil our IR predictions. In fact, the 4-$Q$'s operators in~(\ref{ETC}) with the same symmetry structure as the ones included in~(\ref{pert}) simply renormalize the coupling $f$, whereas those with different symmetry structure (say, vector currents) are expected to be irrelevant. We refer the reader to the quenched QED example~\cite{Bardeen}.}.

It should be clear that the present, strong technicolor approach differs substantially from the strong \textit{extended} technicolor scenarios discussed in~\cite{SETC}\cite{SETC'} and~\cite{SETC''}, where the operators in~(\ref{ETC}) were assumed to contribute to chiral symmetry breaking. On the contrary, here we would like to argue that those operators can be consistently decoupled, and in particular that the enhancement of the chiral condensate $\langle\bar QQ\rangle$ compared to WTC (see below for more details) is a robust physical prediction of TCC (in fact defined for $\Lambda_{ETC}\rightarrow\infty$) .

We will now find an upper bound for the flavor scales $\Lambda_{ETC}^i$ by estimating the energy at which the SM fermions become strongly coupled~\cite{CTC}. We then compare it with the more conservative estimate of $\Lambda_{ETC}^i$ proposed in~\cite{Ratt}, and show that in both cases $(\Lambda_\chi/\Lambda_{ETC}^i)^2\ll1$ for any generation $i=1,2,3$.

The running scaling dimension $\Delta_{\bar QQ}$ in TCC has been computed at leading order in the planar expansion, see~(\ref{beta}), and it should be a reliable estimate down to energies of order $\Lambda_\chi$. The RG flow in the IR depends on the parameter $\epsilon<1$ introduced in Section~\ref{N}. For the sake of illustration, here we will consider the rather optimistic scenario $\epsilon=0$; one can verify that the following discussion is very mildly sensitive to the parameter $\epsilon$ as long as $\epsilon\lesssim O(0.1)$~\cite{CTC} (this latter bound will be motivated in Section~\ref{d}). 

Under these simplifying assumptions we re-write~(\ref{dim}) as:
\ba\label{approx}
\Delta_{\bar QQ}(\mu)=\left\{ \begin{array}{ccc}  1\quad&\mu<e\Lambda_\chi&  \\\\\vspace{2pt} 
2-\frac{1}{\log\left(\mu/\Lambda_\chi\right)}\quad&\mu>e\Lambda_\chi,& \end{array}\right.
\ea
where $e=2.718\dots$ is Neper's number. Now, imposing the boundary condition $\bar y_t(v)=1$, from~(\ref{Ybeta}) we obtain
\ba
\bar y_t(\mu)=\left\{ \begin{array}{ccc}  1\quad&\mu<e\Lambda_\chi& \\\\\vspace{1.5pt} 
\frac{\mu}{e\Lambda_\chi}\frac{1}{\log\left(\mu/\Lambda_\chi\right)}\quad&\mu>e\Lambda_\chi.& \end{array}\right.
\ea
This formula expresses the statement that the chiral condensate $\langle\bar QQ\rangle$ in TCC is enhanced at low energy compared to a WTC model by a factor $\sim e\log\left(\mu/\Lambda_\chi\right)$. A typical SM fermion mass in TCC is hence subject to both a suppression from the flavor scale $\Lambda_{ETC}^i$ and a $\sim e\log\Lambda_{ETC}^i$ enhancement. Because of these two contributions, fermion mass generation in TCC turns out to share some similarities with both the WTC scenario (where approximately $\Delta_{\bar QQ}\gtrsim2$ all the way to $\Lambda_{ETC}$) and the weakly coupled Higgs scenario ($\Delta_{\bar QQ}\sim1$). Specifically, the dependence of a typical SM fermion mass on the ratio $\Lambda_\chi/\Lambda_{ETC}^i$ is relaxed as opposed to a walking dynamics (though enormously enhanced with respect to an ordinary weakly coupled Higgs model); as a result, the flavor scale $\Lambda_{ETC}^i$ in TCC can generally be much higher than in WTC, as we now show.

Requiring that $\bar y_t(\mu)<4\pi$ we find that the flavor scale cannot exceed a value $\Lambda_{NP}^{i=3}\gtrsim\Lambda_{ETC}^{i=3}$ of the order:
\ba
\Lambda_{{NP}}^{i=3}=(150\div200)\times\Lambda_\chi.
\ea
In TCC the top quark becomes strongly coupled to the Higgs sector at a scale a factor $\gtrsim14$ larger than expected in a WTC model.


At scales of $O(\Lambda_{NP}^{i=3})$ the top Yukawa becomes non-perturbative, and our analysis becomes unreliable. We therefore interpret $\Lambda_{NP}^{i=3}$ as an upper bound for the ``flavor scale" $\Lambda_{ETC}^{i=3}$ at which new structures -- involving, at least, top quarks -- might become relevant to reestablish perturbation theory. 
The simplest possible structures we can consider are 4-SM quarks interactions of the type~(\ref{ETC}), with coefficients of order
\ba\label{SM4f}
\gamma_{ab}\sim\frac{\bar y^2(\Lambda)}{\Lambda^2}\equiv \frac{1}{\Lambda_{F}^2}.
\ea
A more conservative estimate of the flavor scale is then found by identifying $\Lambda_{ETC}^i$ with the largest value that $\Lambda_F^i=\Lambda/\bar y_i(\Lambda)\lesssim\Lambda_{ETC}^{i}$ can attain~\cite{Ratt}. Assuming as an example that $\Delta_{\bar QQ}$ is nearly constant, this latter approach gives~\cite{Ratt} 
\ba\label{estimate}
\Lambda_F^{i=3}\sim\Lambda_\chi\frac{(4\pi)^{\frac{2-\Delta_{\bar QQ}}{\Delta_{\bar QQ}-1}}}{\bar y_t(\Lambda_\chi)}.\ea
In WTC the scaling dimension of the composite Higgs satisfies $\Delta_{\bar QQ}\gtrsim2$ in the relevant energy range, and eq.~(\ref{estimate}) tells us that the flavor problem in WTC must be addressed around the weak scale, namely $\Lambda_F^{i=3}\lesssim\Lambda_\chi$. 

In TCC, instead, the flavor scale predicted by the conservative approach of~\cite{Ratt} can be as high as $\Lambda_{F}^{i=3}\approx14\times\Lambda_\chi$~\footnote{For comparison, we mention that this latter value would correspond to having a constant $\Delta_{\bar QQ}=1+\epsilon$ with $\epsilon\sim1/2$ in a conformal TC scenario, see~(\ref{estimate}).}. The flavor scales $\Lambda_F^{i=1,2}$ associated to the light flavors are always enhanced by a factor $m_t/m_{light}\gg1$ compared to $\Lambda_F^{i=3}$, and would be large enough to suppress unwanted FCNC effects involving the first two generations.

The generation of realistic SM fermion masses in TCC can be compatible with FCNC effects only if the flavor physics in TCC has a built-in GIM mechanism. In particular, the 4-SM fermion operators suppressed by the scale~(\ref{estimate}) should involve the third quark generation only or, equivalently, the coefficients of the unavoidable 4-SM fermion interactions involving the light generations in~(\ref{ETC}) should be suppressed by 
higher scales $\Lambda_{F}^{i=1,2}$. See~\cite{PA} for a possible realization of this program. 



In any event, we see that the hierarchy $(\Lambda_\chi/\Lambda_{ETC}^i)^2\ll1$ holds for any generation $i=1,2,3$ in TCC. This implies that the flavor physics in TCC is effectively decoupled from the IR, and suggests that the results presented in the present letter should not be significantly sensitive to the details of the short distance physics.

\subsection{A weakly coupled Higgs: the dilaton\label{d}}

In this subsection we would like to elaborate on the physical significance of $\Delta_{\bar QQ}\sim1$, and see if there exists a parametrically light Higgs boson in the class of theories~(\ref{pert}).

Our claim is that TCC has a residual conformal invariance in the IR, and that the quantity $\epsilon$ defined in~(\ref{eps}) parametrizes the explicit CFT breaking. Similarly to what happens in the Gross-Neveu model, the Higgs boson in TCC is hence a pseudo-Nambu-Goldstone mode of scale invariance. In fact, in any dynamics in which the Higgs operator is weakly coupled the physical Higgs boson should be identified with a (pseudo) dilaton field. The proof is rather straightforward. On the one hand, if the Higgs operator is weakly coupled one sees by direct inspection that its couplings are dictated by the low energy theorems of a spontaneously broken approximate scale invariance~\cite{Ellis}\cite{GGS}. On the other hand, if the Higgs itself is a dilaton -- namely if the order parameter of chiral and scale symmetry breaking coincide -- the IR-free nature of the latter implies $\Delta_{\bar QQ}(0)=1$~\cite{dilaton}. 
Whether or not there exists a choice of parameters in TCC for which the physical Higgs, i.e. the pseudo dilaton, can be made parametrically lighter than the other hadrons is an interesting issue we would like to address in the following. Note that this is tantamount to asking whether or not there exists a small parameter controlling the explicit CFT breaking in TCC.

Scale invariance cannot be an exact symmetry in models of dynamical symmetry breaking. Yet, it should be possible to estimate the impact of the scale anomaly on the Higgs physics by studying the effect of the scale anomaly on the dilaton couplings. For definiteness we will focus on the critical theory with $\Delta=2$ in~(\ref{beta}); in the general case $\Delta\neq2$ our strong technicolor theory~(\ref{pert}) presents a hard breaking of conformal invariance and the discussion should be adjusted accordingly.

The model~(\ref{pert}) with $\Delta=2$ is classically scale invariant, but at the quantum level this is no more true. For $f\neq0$ the Ward identity of dilatation invariance is anomalous, $\partial_\mu {\cal D}^\mu={\cal A}$, with an anomaly given by~\footnote{It is easy to check the reliability of the results~(\ref{beta}) by verifying that the anomalous dimension of the matrix element $\langle0|{\cal A}|0\rangle$ vanishes, as it should. Note also that $\langle{\cal A}\rangle=O(N^2)$ is compatible with the equivalence ${\cal A}=\frac{\beta_\lambda}{4\lambda}F_{\mu\nu}^2$ implied by the conjecture proposed in Section~\ref{conj}.}
\ba
{\cal A}=\mu\frac{d f}{d\mu}(\bar QQ)(\bar QQ).
\ea
If there exists a limit in which the scale anomaly may be considered ''small" as compared to the current conservation (we will clarify this statement shortly) then the generation of a non-trivial chiral condensate in the model~(\ref{pert}) would imply a spontaneous breaking of an approximate conformal symmetry. In this limit the longitudinal excitation of the order parameter, i.e. the physical Higgs boson, would be a \emph{light} pseudo-dilaton.


Let us then view the parameter $\epsilon$ defined in~(\ref{eps}) as a measure of the scale symmetry breaking. For $\epsilon=0$ there is no scale anomaly: both chiral and scale symmetries are linearly realized, in particular $\Lambda_\chi=0$. For $\epsilon\neq0$ chiral symmetry breaking and confinement take place, and the dilaton/Higgs mass squared should approximately read
\ba\label{PCDC}
m_\sigma^2=O(\epsilon)\Lambda_\chi^2,
\ea
where $\epsilon$ is controlled by the parameters $N,N_f,\bar f(\Lambda)$ (at leading order in the planar limit $\epsilon=\epsilon(N_f,\bar f)$). By symmetry arguments, it follows that in a model in which the condition~(\ref{PCDC}) is satisfied for some parameter $\epsilon$, the renormalizable couplings of the Higgs/dilaton deviate from those of a fundamental Higgs boson -- equivalently, from those of an exactly massless dilaton with decay constant $f_D=v$ -- by an amount $O(\epsilon)$. Such a conclusion agrees with the results of~\cite{CTC}\cite{SILH}: a departure~(\ref{eps}) from the fundamental Higgs condition $\Delta_{\bar QQ}=1$ implies corrections $O(\epsilon)$ in the couplings of the Higgs.

We were not able to identify any small parameter $\epsilon$ in the theory~(\ref{pert}). Indeed, the NJL analogy suggests that there exists no choice of the external parameters such that~(\ref{PCDC}) with an $\epsilon\ll1$ holds~\footnote{The external parameters $(N_f,\bar f)$ in~(\ref{pert}) would be replaced by $(d,\bar f)$ in the NJL model -- with $d$ the space-time dimension -- and the critical condition $N_f=N_f^c$ would be $d=2$~\cite{Miransky}.}. Specifically, a quantitative analogy with the NJL model gives $\epsilon=O(0.1)$ independently of the value of the external parameters. Furthermore, viewing $\bar f$ as a measure of the effective mass of the quark bilinear, and recalling that $\bar f(\mu\sim\Lambda_\chi)=O(1)$, one would get $m_\sigma^2=O(\Lambda_\chi^2)$ and again conclude that the Higgs boson in TCC is not parametrically lighter than the dynamical scale. 

The bottom line is that, even though an accidental suppression of $\epsilon$ cannot be excluded a priori, we expect that the explicit CFT breaking in TCC is measured by $\epsilon=O(0.1)$, very much like in the NJL model. This is a slightly bigger value than predicted in a truly weakly coupled scalar sector in which case, say, $\epsilon$ would be of the order of a loop factor. Nevertheless, it is sensible to refer to the Higgs field of TCC as \textit{weakly coupled} because its conformal weight is much closer to that of a free Higgs field than to its engineering dimension. The consequences of this fact on flavor physics have been discussed in Section~\ref{fla}. The implications on the EW precision measurements are difficult to estimate; see~\cite{Q} for a holographic approach.






\section{\label{conj}Asymptotic (non)freedom in~gauge~theories~?!}

In this section we will suggest a physical interpretation of the path integral~(\ref{pert}).

It is known that non-abelian gauge theories posses a conformal window, i.e. a finite range in the number $N_f$ of massless flavors in which the UV free theory becomes conformal in the IR. However, little is known about the actual critical number of massless flavors $N_f^c$ below which conformality is lost. Most of our knowledge comes from the study of unsystematic truncations of the \SDE or the analogy with the supersymmetric example. Yet, if chiral symmetry breaking is responsible for the loss of conformality at the lower end of the conformal window (for $N_f\leq N_f^c$), then the supersymmetric analogy cannot be a guide to the physics: the study of the \SDE is much more appropriate~\cite{Vecchi}.

\begin{figure}
\begin{center}
\includegraphics[width=120mm, height=70mm
]{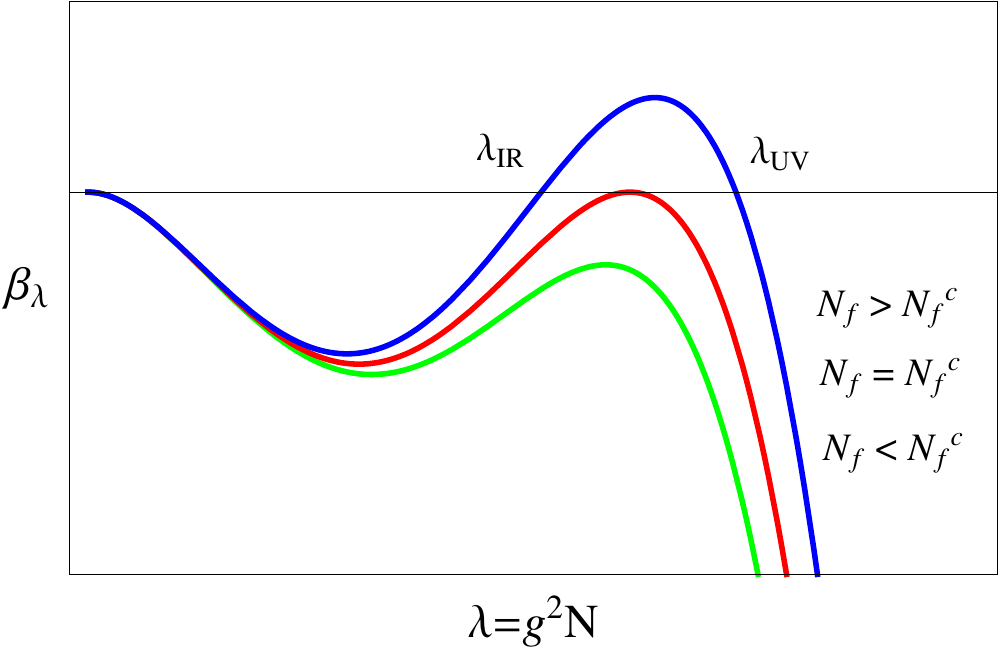}
\caption{\small  Conjectured beta function for the 't Hooft coupling $\lambda$ of a non-supersymmetric strong dynamics as a function of $\lambda$ for three different values of $N_f$ in the vicinity of the critical value $N_f^c$, and below the bound at which asymptotic freedom is lost (see also footnote~\ref{foot}). In the lower line $N_f<N_f^c$, in the middle line $N_f=N_f^c$, and in the upper line $N_f>N_f^c$. The first case is the one advocated in the WTC framework; in this section we will focus on the other two cases. The conjecture predicts the existence of a non-trivial UV fixed point $\lambda_{UV}$ in addition to the IR fixed point $\lambda_{IR}\leq\lambda_{UV}$. The fixed points are expected to merge $\lambda_{IR}=\lambda_{UV}=\lambda_c$ at the lower end of the conformal window, i.e. when $N_f=N_f^c$, and disappear when $N_f<N_f^c$. \label{fig1}}
\end{center}
\end{figure}

Recently, the authors of~\cite{CL} conjectured that the beta function for the 't Hooft coupling $\lambda=g^2N$ of non-supersymmetric non-abelian gauge theories varies as a function of $N_f$ as indicated in Fig.~\ref{fig1} (for $N_f$ close to the critical value $N_f^c$ and still in the UV-free phase). This conjecture implies the existence of a non-trivial UV fixed point $\lambda_{UV}$ in addition to the IR fixed point $\lambda_{IR}$ characterizing the conformal window~\footnote{\label{foot}Beta functions are unphysical (scheme-dependent). Yet, there are a number of unambiguous properties of the associated physical system that can be extracted from them, such as the existence of fixed points separating different phases, and the relative critical exponents. Fig.\ref{fig1} is meant to pictorially express two physical statements: 1) the phase $\lambda>\lambda_{UV}$ has chiral symmetry breaking and confinement; 2) when $N_f<N_f^c$ the IR-conformal and the strong phases merge and chiral symmetry breaking and confinement are realized in the UV-free phase.}. A confirmation of this picture certainly requires a careful treatment of the chiral limit on the lattice~\cite{Hasenfratz}.

What renders the above conjecture appealing is the fact that it essentially captures all of the main properties that have been extracted from studies of the \SDE\cite{Vecchi}. In particular, the beta function in Fig.~\ref{fig1} encodes the fact that chiral symmetry breaking signals the lower end of the conformal window at $N_f=N_f^c$, and thereby explains the loss of conformality as a conformal phase transition~\cite{Miransky} at some critical scaling dimension $\Delta_c$ for the fermion condensate. The value $\Delta_c=2$ has been speculated to be the true, non-perturbative signal for chiral symmetry breaking in~\cite{CG}.

Our primary observation is that, if the picture emerging from Fig.~\ref{fig1} is correct -- and ultimately if the \SDE approach actually captures the relevant features of non-supersymmetric non-abelian gauge theories --, then there exists a region in the coupling space of non-abelian gauge theories in which the theory is asymptotically \textit{non-free} and yet spontaneously breaks the chiral symmetry. The asymptotically non-free regime would be found for strong renormalized 't Hooft couplings $\lambda$ bigger than the new, non-trivial UV fixed point $\lambda_{UV}\geq\lambda_{IR}>0$ depicted in Fig.~\ref{fig1}. In this section we would like to address the physical implications of such a regime.

\subsection{An effective approach, and~a~conjecture}

In order to address the physical relevance of the asymptotically non-free branch of Fig.~\ref{fig1}, we need a model for the strong ($\lambda\geq\lambda_{IR}$) technicolor theory. There are at least two approaches to this program. The first consists in describing the strong branch in terms of the original non-abelian gauge theory action. This approach requires a nonperturbative tool. The second approach is more adequate to our purposes, and consists in formulating a theory for the strong branch in terms of a \textit{dual} field theory defined at $\lambda_{UV}$. An effective formulation of such a dynamics is obtained by including on the top of the CFT defined by the TC theory at the IR fixed point $\lambda_{IR}$ all the operators ${\cal O}_i$ that are relevant at the non-trivial UV fixed point
. The formal description of the asymptotically non-free dynamics would hence be given in terms of the path integral~\cite{Vecchi}
\ba\label{theory}
\langle e^{i\int \sum_if_i{\cal O}_i}\rangle_{CFT}.
\ea
In the above expression, the CFT is defined by the correlators of the non-abelian theory at the IR fixed point $\lambda=\lambda_{IR}$, the ${\cal O}_i$'s are local operators that become relevant at $\lambda_{UV}$, and $f_i$ are suitable couplings for the CFT perturbations. The $f_i$'s represent the \textit{only} couplings in our dual (effective) description. 

In principle, there exists a neat way to identify the set of local operators relevant to our analysis, at least for $\lambda_{UV}\sim\lambda_{IR}$ (i.e. for $N_f\sim N_f^c$). One defines the asymptotically free theory on the lattice for a number of flavors $N_f\geq N_f^c$, and then studies the RG evolution of the local operators. At the IR fixed point $\lambda=\lambda_{IR}$ one identifies a set of operators with scaling dimension $\Delta\leq4$; by continuity, we expect these dimensions to be arbitrarily close to the UV dimensions of the corresponding operators defined at the UV fixed point $\lambda_{UV}$ when $N_f$ is arbitrarily close to $N_f^c$, i.e. when the upper curve merges the middle curve in Fig.~\ref{fig1}. The set of operators with IR dimension $\Delta\leq4$ represents the complete set of (UV) relevant deformations defining the asymptotically non-free branch at $\lambda=\lambda_{UV}(\sim\lambda_{IR})$.


Clearly, there is no known analytical method which can unambiguously determine such a set of operators, mainly because of the intrinsic non-perturbative nature of the problem. What we certainly know is that the operators ${\cal O}_i$ defined in~(\ref{theory}) must be flavor symmetric and irrelevant at $\lambda_{IR}$ (the latter requirement follows from the observation that the IR physics of the branch $\lambda_{IR}<\lambda<\lambda_{UV}$ should be governed by the very same IR fixed point found in the asymptotically free branch $\lambda<\lambda_{IR}$). A possible hint on the class of operators we should take into account comes from the study of the \SDE and the analogy with quenched QED at large coupling~\cite{Vecchi}. These considerations suggest that the flavor-singlet 4-fermion contact term (gauge indices are contracted inside the parenthesis)
\ba\label{4ferm}
(\bar QQ)(\bar QQ)
\ea
is one of the relevant operators ${\cal O}_i$ defining the strong branch. If~(\ref{4ferm}) were the \textit{only} deformation, the path integral~(\ref{theory}) would simplify to the (gauged) Nambu-Jona Lasinio dynamics~(\ref{pert}). Because in this latter case, as discussed in Section~\ref{TCC}, no additional operators would be strictly required at leading order in the planar limit, we are tempted to \textit{conjecture}~(\ref{pert}) \textit{to be the large N dual, effective description of the strongly coupled branch of} Fig.~\ref{fig1}. Similarly, the quenched QED model for chiral symmetry breaking -- the abelian version of~(\ref{pert}) -- would be an effective description of the strong branch of abelian gauge theories, as suggested in~\cite{Bardeen}. Let us now see if our interpretation is sensible.




Because the 4-fermion operator~(\ref{4ferm}) is understood to be generated by the strong dynamics, we will view $f$ in~(\ref{pert}) as an unknown function of the 't Hooft coupling, i.e. $\bar f=\bar f(\lambda)$, very much like the parameters of the chiral lagrangian may be thought of as functions of the QCD coupling. We will present an explicit mapping $\lambda\rightarrow\bar f$ shortly. For the moment we emphasize that if we insist with this interpretation, we should expect the beta function of $\bar f$ to ''encode" the running of $\lambda$ in the strong branch. This would allow us to make a number of non-trivial checks of the consistency of the conjecture.


Let us hence assume that the beta function of the 't Hooft coupling in the regime $\lambda>\lambda_{IR}$ is given at leading $1/N$ order by~(\ref{beta})~\footnote{Note that for $\bar f<0$ there appears an UV Landau pole at finite energy and the 4-fermion operator becomes trivial: our formalism relates the region $\bar f<0$ to the asymptotically free phase $\lambda<\lambda_{IR}$ -- in which the 4-fermion operator must be switched off and our dual description is expected to fail -- whereas the region $\bar f\geq0$ to the strong branch $\lambda>\lambda_{IR}$.}
\ba\label{map}
\beta_\lambda=\frac{\beta_{\bar f}}{\bar f'}=\frac{-\bar f^2+(2\Delta-4)\bar f}{\bar f'},~~~~~~~~~~~\bar f'=\frac{d\bar f}{d\lambda}.
\ea
Now, the beta functions~(\ref{beta}) and~(\ref{map}) tell us that the coupling $\bar f$ (i.e. $\lambda$) develops a trivial IR fixed point $\bar f_{IR}=0$ and an UV fixed point $\bar f_{UV}=2\Delta-4$. The trivial fixed point was anticipated, and reflects the statement that the 4-fermion perturbation is actually irrelevant at $\lambda_{IR}$, i.e. the IR physics in the phase $0=\bar f_{IR}\leq\bar f\leq\bar f_{UV}$ is governed by the undeformed CFT. The UV fixed point should be associated to the occurrence of an UV fixed point $\lambda_{UV}$ for the 't Hooft coupling, as dictated by the relation $\bar f_{UV}=\bar f(\lambda_{UV})$. The fixed points $\lambda_{IR,UV}$ are thus correctly reproduced by our formalism. What about chiral symmetry breaking? 

Chiral symmetry breaking manifests itself in the dual theory as an instability of the description~(\ref{pert}), like in the NJL model or in quenched QED. This occurs in the regime $\bar f>\bar f_{UV}$, where the beta function in~(\ref{beta}) develops an IR Landau pole. In the gauge theory language chiral symmetry breaking is then predicted to occur in the phase $\lambda>\lambda_{UV}$, as suggested by Fig.\ref{fig1}.

\begin{figure}
\begin{center}
\includegraphics[width=120mm, height=70mm]{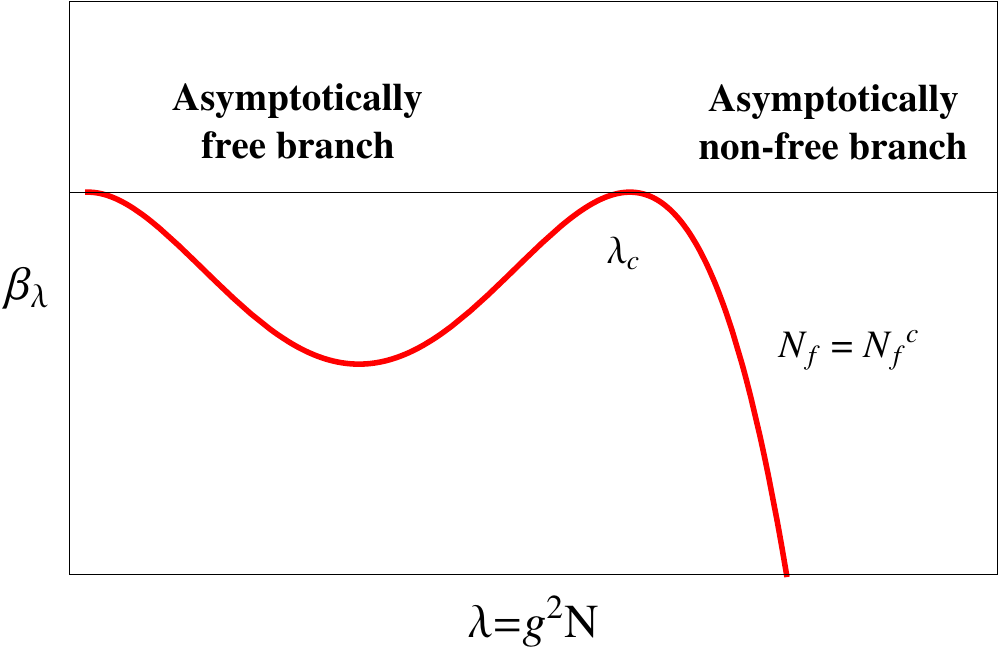}
\caption{\small  Conjectured beta function for the 't Hooft coupling $\lambda$ of a non-supersymmetric strong dynamics as a function of $\lambda$ for a critical number of fermions $N_f=N_f^c$ (see also footnote~\ref{foot}). The non-trivial zero $\lambda_c$ resulting from the merger of $\lambda_{IR}$ and $\lambda_{UV}$ in Fig.~\ref{fig1} is an IR fixed point for the asymptotically free branch $\lambda<\lambda_c$, and an UV fixed point for the strong branch $\lambda>\lambda_c$. The critical dynamics formulated in the strong branch $\lambda\geq\lambda_c$ is asymptotically non-free, natural, and spontaneously breaks the chiral symmetry. We conjectured TCC, see~(\ref{pert}), to be an effective description of this dynamics. \label{fig2}}
\end{center}
\end{figure}

As $\bar f_{UV}\rightarrow0$ the techni-quark bilinear acquires scaling dimension $\Delta=2$ (see~(\ref{beta})) and the dual theory undergoes a conformal phase transition~\cite{Miransky}. In this limit chiral symmetry breaks down in the phase $\bar f>0$, and the scale of the order parameter is (see~(\ref{LL}) with $\delta<\bar f$)
\ba\label{Lambda}
\Lambda_\chi=\Lambda\, e^{-\frac{1}{\bar f(\Lambda)}}.
\ea
We recognize the exponential relation between the renormalized coupling and the dynamical scale characterizing models with a natural hierarchy of scales: the theory (\ref{pert}) is natural in the limit $\bar f_{IR}=\bar f_{UV}$ (i.e. $\Delta=2$) in which the CFT perturbation becomes marginal (marginally relevant indeed). Recalling that $\bar f=\bar f(\lambda)$ we may rephrase this physics as follows. As the number of flavors decreases $N_f\rightarrow N_f^c+0^+$ the fixed points $\lambda_{IR}$ and $\lambda_{UV}$ merge. In this limit the chiral symmetry is broken in the strong phase $\lambda>\lambda_{UV}=\lambda_{IR}$ (i.e. $\bar f>0$) when the techni-quark bilinear reaches the critical dimension $\Delta(\lambda_{IR})=2$, consistently with the claim of~\cite{CG}. Chiral symmetry breaking becomes visible to the asymptotically free branch only for $N_f<N_f^c$, when our dual description breaks down. The critical theory defined at $N_f=N_f^c$ is depicted in Fig.~\ref{fig2}.

The emerging picture matches remarkably well with the physics of chiral symmetry breaking in non-supersymmetric non-abelian gauge theories extracted from studies of the \SDE\cite{SDE}\cite{ATW} (see also~\cite{CG} for a pedagogical presentation of the analysis). In the ladder approximation, the dynamical fermion mass $\Sigma(q=0)$ obtained in those studies takes the form
\ba\label{SDE}
\Sigma=\Lambda\, e^{-\pi/\sqrt{\frac{\lambda}{\lambda_c}-1}},\quad~~~~~\lambda_c=\frac{4\pi^2N}{3C_2(R)}
\ea
with $\Lambda$ an UV cutoff and $C_2(R)$ the quadratic Casimir for the fermion representation $R$. Consistency of the solution~(\ref{SDE}) requires $\lambda>\lambda_c$ and a very slowly running coupling $\lambda(\Lambda)\sim\lambda_c$. The RG equation for $\lambda(\Lambda)$ is obtained by demanding that $\Sigma$ does not depend on the cutoff; this gives the so called Miransky scaling~\cite{SDE}:
\ba\label{mir}
\beta_\lambda=-\frac{2}{\pi}\lambda_c\left(\frac{\lambda}{\lambda_c}-1\right)^{3/2}.
\ea
We can then deduce two important physical implications of the \SD equation approach to chiral symmetry breaking. First, the critical coupling $\lambda_c$ of eq.~(\ref{SDE}) must be interpreted as an UV attractive fixed point. This in turn justifies the slowly varying condition $\lambda\sim\lambda_c$ assumed in these studies. Second, at $\lambda=\lambda_c$ the non-abelian gauge theory is defined by marginal deformations, i.e. the beta function $\beta_\lambda$ has a flat tangent at $\lambda_c$. This explains the occurrence of a conformal phase transition at $\lambda=\lambda_c$~\cite{Miransky}.

These results are nicely reproduced by Fig.~\ref{fig2} and the model~(\ref{pert}) if we interpret $\lambda_c$ as the critical coupling at which $\lambda_c=\lambda_{IR}(N_f=N_f^c)=\lambda_{UV}(N_f=N_f^c)$. Specifically, identifying the dynamical mass~(\ref{SDE}) with the scale $\Lambda_\chi$ at which the coupling of the dual description blows up, see~(\ref{Lambda}), we have
\ba\label{mapping}
\bar f(\lambda)=\frac{\sqrt{\frac{\lambda}{\lambda_c}-1}}{\pi}+\dots
\ea
where the dots stand for subleading corrections in $\lambda-\lambda_c$. Plugging this mapping into~(\ref{map}) (recall that $\Delta=2$ at the critical point $N_f=N_f^c$) we consistently find~(\ref{mir}). We thus see that our conjecture suggests to view the Miransky scaling~(\ref{mir}) as the running flow equation for the 't Hooft coupling of non-abelian gauge theories in the overcritical phase $\lambda>\lambda_c$ of Fig.~\ref{fig2}, with $\lambda_c$ representing an UV attractive fixed point. The non-triviality of the fixed point $\lambda_c$ would in turn clarify the concerns raised in~\cite{CG} on the applicability of the OPE to the strong phase $\lambda>\lambda_c$.

The above observations are certainly not enough to prove our conjecture that~(\ref{pert}) is indeed capable of describing the physics of chiral symmetry breaking in non-abelian gauge theories. Yet, we believe that the arguments presented above are at least suggestive.


\section{Conclusions}
\label{Con}

We presented an asymptotically \textit{non-free}, natural model for dynamical electro-weak symmetry breaking -- technicolor at criticality (TCC). The theory has been defined in terms of the path integral~(\ref{pert}), and may be seen as a generalization of the quenched QED model for chiral symmetry breaking. 
A striking feature of our model is that it is treatable in a leading planar expansion.


The scaling dimension of the techni-quark bilinear $\bar QQ$ in TCC is found to lie in the range~(\ref{gamma}). This property is not in conflict with naturalness. In particular, we have seen that the low energy physics of TCC depends at most logarithmically on the RG scale.

The UV condition $\Delta_{\bar QQ}=2$ is realized if the number of massless flavors in the underlying CFT is chosen to be at a critical value,  and ensures that the dynamical scale $\Lambda_\chi=O(1)$ TeV be naturally smaller than the UV cutoff $\Lambda$. The IR condition $\Delta_{\bar QQ}\sim1$ implies that the long distance physics of TCC is weakly coupled. This latter fact has important consequences on flavor physics, and potentially on the EW precision observables as well.

The relation $\Delta_{\bar QQ}\sim1$ is crucially linked to the existence of an approximate, nonlinearly realized scale invariance, and suggests to interpret the Higgs boson in TCC as a pseudo-dilaton. In this respect, the physics responsible for the emergence of a weakly coupled Higgs sector in the IR regime of~(\ref{pert}) is analogous to the one invoked in composite Higgs models; the main difference between the two realizations is that in the latter case the Higgs is a pseudo-\NG mode of some \textit{exact global} symmetries of the strong dynamics, whereas in TCC the Higgs appears as a pseudo-\NG mode of an \textit{approximate dilatation} invariance of the strong sector.

The beyond the standard model phenomenology of TCC is characterized by the presence of sharp resonances ($\Gamma/M\propto1/N^2$) of various spins and masses $M=O(\Lambda_\chi)$. Despite the existence of a weakly coupled Higgs sector, the low energy physics of TCC cannot be captured by a linear sigma model: the Higgs boson is not parametrically lighter than the other hadrons. Yet, the approximate conformal invariance can help us guessing the form of the Higgs effective action. Indeed, the breaking of conformality should be encoded in the \textit{logarithmic} running of the coupling $\bar f$, and the leading effective action for the Higgs field $H\propto\bar QQ$ should hence be of the form~\cite{Miransky}\cite{Rat}\cite{GGS}
\ba\label{pot}
f_D^2\, H^\dagger D^2H-|H|^4\, {\cal V}\left(\log \frac{|H|}{\Lambda}\right)
,
\ea 
where 
${\cal V}=O(N^2)=f_D^2=v^2(1+O(\epsilon))$ in our planar expansion. When the CFT deformation is switched on the vacuum condition becomes $\log\langle |H|\rangle/\Lambda=O(1)$, which is our~(\ref{LL}). There is no small parameter in the potential apart from $1/N^2$. Accordingly, the Higgs boson is expected to have a mass $m_\sigma=O(\Lambda_\chi)$ comparable to the other hadrons, and the nontrivial $H$-vertices are suppressed by powers of $1/N^2$.


Large $N$ models satisfying~(\ref{gamma}) must have a relatively low flavor scale if they want to avoid power-law sensitivity on the UV cutoff. In the specific case of TCC, we saw that the top physics becomes strongly coupled at energies $\sim(150\div200)\times\Lambda_\chi$. Using the rather conservative approach adopted in~\cite{Ratt}, such a high non-perturbative scale translates into a flavor scale for the third SM quark generation of order $\Lambda_{ETC}^{i=3}\approx14\times\Lambda_\chi$. This prediction will hopefully be testable at future collider experiments. 




We proposed to interpret~(\ref{pert}) as a dual, effective description of the strongly coupled phase of non-supersymmetric non-abelian gauge theories conjectured in~\cite{CL}\cite{Vecchi}. Due to the IR robustness of our analysis of the theory~(\ref{pert}), if this latter interpretation is correct TCC may as well capture the long distance physics of more conventional (asymptotically-free) scenarios for dynamical breaking. We presented a few suggestive arguments in favor of this conjecture relying on a novel interpretation of the results extracted from the \SD equation approach to chiral symmetry breaking, and in particular of the Miransky scaling~\footnote{After the first version of this paper was submitted on the ArXiv the authors of~\cite{HY} proposed a similar interpretation of the Miransky scaling. Their conclusions on the absence of a parametrically light dilaton in non-abelian gauge theories (the absence of a controllable small $\epsilon$ in our language) also agree with ours.}.

\acknowledgments

This work is dedicated to Michele Turrini, a dear friend.

We are grateful to Michael L. Graesser for valuable discussions. This work has been supported by the U.S. Department of Energy at Los Alamos National Laboratory under Contract No. DE-AC52-06NA25396.


 \end{document}